\documentclass{article}

\usepackage{graphicx}
\usepackage{amsmath}

\def\giorno{29 February 2008}

\def\a{\alpha}
\def\b{\beta}

\def\ga{\gamma}
\def\de{\delta}   
\def\eps{\varepsilon}
\def\la{\lambda}
\def\s{\sigma}

\def\om{\omega}

\def\R{{\bf R}}

\def\<{\langle}
\def\>{\rangle}

\def\({\left(}
\def\){\right)}
\def\[{\left[}
\def\]{\right]}
\def\=#1{\bar #1}
\def\~#1{\widetilde #1}
\def\.#1{\dot #1}
\def\^#1{\widehat #1}
\def\"#1{\ddot #1}

\def\salta#1{{}}

\def\beq{\begin{equation}}
\def\eeq{\end{equation}}
\def\feq{\end{equation}}

\begin{document}

\title{A mean-field version of the SSB model for X-chromosome
inactivation}

\author{Giuseppe Gaeta \\ Dipartimento di Matematica, Universit\`a di
Milano, \\ via Saldini 50, 20133 Milano (Italy) \\
{\tt gaeta@mat.unimi.it}}

\label{firstpage}
 \date{{\tt \giorno}}

\maketitle



\begin{abstract}
\noindent  Nicodemi and Prisco recently proposed a model for
X-chromosome inactivation in mammals, explaining this phenomenon
in terms of a spontaneous symmetry-breaking mechanism [{\it Phys.
Rev. Lett.} 99 (2007), 108104]. Here we provide a mean-field
version of their model.
\end{abstract}

\section{Introduction}

The nucleus of female mammals embryo cells contain two X
chromosomes; one of these has to be inactivated in order to
equalize the dosage of X genes product with respect to males, and
this inactivation is crucial to survival \cite{CYZB,LKP,CTTB}.
Moreover, the X chromosome is conjectured to play a key role in
the arising of certain types of cancer \cite{JNCI1,JNCI2}.

The mechanism by which one of the two X chromosomes is inactivated
is poorly understood, despite extensive work on this problem
\cite{CYZB,LKP,CTTB}; it is known that co-localization of
chromosomes at the time X-chromosome inactivation (XCI) is
established plays a key role \cite{NTL,NCB,Xclose}.

Very recently, Nicodemi and Prisco \cite{NP} proposed a model to
explain XCI on the basis of a spontaneous symmetry-breaking
mechanism, related to the classical Ising model of Statistical
Mechanics, in a dynamical stochastic model of the single-agent
type.

Their idea is that blocking factor (BF) molecules diffuse in the
cell and can bind randomly to one or the other of the
X-chromosomes; however, by a collective effect, the affinity to a
chromosome increases when this already binds to other BF
molecules. The net effect is that once one of the two chromosomes,
by the effect of a fluctuation, binds to a larger number of BF
molecules, it will bind to more and more of these -- at a rate
higher than the other chromosome (note that the BF molecules could
have a greater probability to escape when more of them bind to the
chromosome; this effect should be dominated by the growth in
affinity in order for the collective effect devised by Nicodemi
and Prisco to set in). In the whole, as a result of diffusive
behavior, fluctuation and the affinity-enhancing collective
effect, a net current will be established leading to BF
concentration on one of the two X-chromosomes.

The model can also account for the relevance of co-localization at
the establishment of XCI; in facts, if the two X-chromosomes are
too far away, the diffusive behavior will prevail, and the BF
molecules escaping from the less populated X-chromosome will have
a very low probability to reach the other one.

The purpose of this short note is to provide a deterministic
(simplified) version of the Nicodemi-Prisco symmetry-breaking XCI
model, considering average exchanges of blocking factors between
each of the two X-chromosomes and the cell fluid environment. Thus
our model will amount to a mean-field version of the
Nicodemi-Prisco model; we will formulate it first in terms of
transition probabilities between states for a single BF molecule,
and then pass to a formulation in terms of a first order system of
ODEs, i.e. a Dynamical System \cite{Gle,GuH,Ver}, following a
procedure which is standard in Physics or in Quantitative Biology
\cite{Mur,ElG}.

Here we will consider the fluid environment as a homogeneous
reservoir, i.e. overlook the effects connected with the spatial
localization of the X-chromosomes; thus the model will be able to
predict the symmetry breaking leading to inactivation of one of
the X-chromosomes, but (in its simplest form, presented here) not
its dependence on co-localization of the chromosomes. An extension
of the model aiming at including co-localization effects will be
developed in forthcoming work.

Similarly, here we will only deal with average quantities and
flows; fluctuations can be included within this kind of modeling
by promoting the ODEs we obtain to {\it stochastic differential
equations} \cite{Gue,Oks,vKam}, as briefly discussed below.

\section{The model}
\label{sec:model}

We will consider a fixed total amount ($n$ units) of BF molecules;
these can be in three mutually exclusive states: they can bind to
one (state 1) or the other (state 2) of the two X-chromosomes, or
be diffusing in the cell environment (state 3).  We denote the
number of units binding to the first X-chromosome at time $t$ as
$\^X(t)$, that of units binding to the second X-chromosome at time
$t$ as $\^Y(t)$, and that of units diffusing in the cell fluid at
time $t$ as $\^Z(t)$. We will refer to $x(t)$ and $y(t)$ as the
{\it occupation numbers} for the two X-chromosomes (or more
precisely, for the Xic of the chromosomes).  Obviously,
\beq\label{totnumber} \^X(t) + \^Y(t) + \^Z(t) \ = \ N \ . \eeq

\subsection{Transition probabilities for a single BF molecule}
\label{sec:mod:TP}

Let us consider the probabilities $P(i \to j, \de t)$ for
transitions between these states -- in particular, transition from
state $i$ to state $j$ -- for each BF molecule (referred to as a
``particle'' in the following) in a time interval of length $\de
t$. In the following, we will use a simplified notation, and write
\beq\label{probshort} p_{ij} \ := \ P(i \to j , \de t) \ . \eeq

We assume that for $\de t$ sufficiently small, the probability a
particle initially in state 3 will end up being in state 1 or 2
(i.e. bound to one of the two X-chromosomes) is \beq\label{prob1}
p_{31} \simeq  \a (\^X)  \de t \, + \, o (\de t) \ ; \ \ p_{32}
\simeq \a (\^Y)  \de t \, + \, o (\de t) \ . \eeq That is, it will
be proportional to the length of the time interval (but
independent of time $t$ itself) via a function $\a (x)$ or $\a
(y)$ representing the affinity with the concerned X-chromosome and
depending on the number of particles already binding to it.

Similarly, the probability that a particle initially binding to
one of the X-chromosomes will escape from it in a time interval
$\de t$ will be given by \beq\label{prob2} p_{13} \simeq
\b(\^X,\^Z) \de t \, + \, o (\de t) \ ; \ \ p_{23} \simeq
\b(\^Y,\^Z) \de t \, + \, o (\de t) \ . \eeq
 That is, the escape probability will be
proportional to the length of the time interval via a function of
the number of particles binding to the concerned X-chromosome and
of the number of particles fluctuating in the fluid environment
(it is natural to assume these should enter only through their
density $z = Z/N$).

As for direct transitions from one to the other of the two
X-chromosomes, these will be impossible on account of their
non-zero space separation (the particle will have to diffuse
through the fluid to do that), so we will set \beq\label{prob3}
p_{12} \ = \ o (\de t) \ ; \ \ p_{21} \ = \ o (\de t) \ . \eeq

Needless to say, the probabilities of permanence in a given state
are obtained simply requiring the conservation of particles, i.e.
that the sum of probabilities for all transitions from a given
state to all possible state is one (Markov property); thus
\beq\label{prob4}
\begin{array}{rl}
p_{11} \ =&  1 - p_{12} - p_{13}  =  1 - \b(\^X,\^Z) \de t + o (\de t) \ , \\
p_{22} \ =&  1 - p_{21} - p_{23}  =  1 - \b(\^Y,\^Z) \de t + o (\de t) \ , \\
p_{33} \ =&  1 - p_{31} - p_{32}  =  1 - [\a(\^X) + \a(\^Y)] \de t
+ o (\de t) \ .
\end{array} \eeq

\subsection{Evolution equations for occupation numbers}

Let now $\{ X(t),Y(t),Z(t)\}$ be the average occupation numbers
for states $\{1,2,3\}$ at time $t$; we look at the average
occupation numbers at time $t + \de t$. Note $Z(t)$ can be
expressed using \eqref{totnumber}, hence we need only two
equations. These will be given by \beq\label{eveq1a}
\begin{array}{rl} X (t+ \de t)  =&  X(t)  - p_{13} X(t) + p_{31} Z(t)  \ , \\
 Y (t+ \de t)  =& Y(t)  - p_{23} Y(t) + p_{32} Z(t)
\ .
\end{array} \eeq

Using the expressions given above for $p_{ij}$ -- and omitting $o
(\de t)$ terms -- the \eqref{eveq1a} yield \beq\label{eveq1}
\begin{array}{rl}
 X (t+ \de t)  =& X(t) + [ \a(X(t)) \, Z(t) \, - \, \b(X(t),Z(t)) \, X(t) ] \de t  \\
 Y (t+ \de t)  =& Y(t) + [ \a(Y(t)) \, Z(t) \, - \, \b(Y(t),Z(t)) \, Y(t) ] \de t  
\ .
\end{array} \eeq

Dividing by $\de t$ and letting $\de t \to 0$, eq.\eqref{eveq1}
yields a system of ODEs describing the evolution of the average
occupation numbers for the three states (in which we make explicit
the expression of $Z(t)$ implied by the conservation law
\eqref{totnumber}): \beq\label{eveq3}
\begin{array}{rl}
d X / dt  =&  \a(X) \, (N - X - Y) \ - \ \b(X,N-X-Y) \, X \ , \\
d Y / dt  =&  \a(Y) \, (N - X - Y) \ - \ \b(Y,N-X-Y) \, Y \ .
\end{array} \eeq

Passing to consider the densities $x = X/N$ and $y = Y/N$, and
introducing the functions $a$ and $b$ defined through \beq a (w)
:= \a (N w) \ , \ \ b(w_1,w_2) := \b (N w_1,N w_2) \ , \eeq we get
in the end the equations (symmetric under the exchange of $x$ and
$y$) \beq\label{eveq4}
\begin{array}{l}
d x / dt \ = \ a(x) \, (1-x-y) \ - \ b(x,1-x-y) \, x \ , \\
d y / dt \ = \ a(y) \, (1-x-y) \ - \ b(y,1-x-y) \, y \ .
\end{array} \eeq

\subsection{Minimal model}

So far we have worked in completely general terms within the
three-states description of the system. In order to have a
specific model, we should choose concrete forms for the functions
$\a(W)$ and $\b (W)$ -- and hence of the $a(w)$ and $b(w)$ --
describing the probability of capture by and escape from
chromosomes in function of the associated occupation numbers $X,Y$
or densities $x,y$; this will yield a specific expression for the
evolution equations \eqref{eveq3} and \eqref{eveq4}.

In their work, Nicodemi and Prisco \cite{NP} argued that the
essential feature of their model for XCI is the collective
phenomenon enhancing affinity and hence the probability of capture
by X-chromosomes with a substantial number of binding particles.

Thus the function $\a (W)$ should go to a finite limit $\a_0$ for
$W \to 0$ (the BF have a non-zero affinity with the chromosomes
even when no BF molecules are binding to chromosomes, and more
generally when the collective behavior has not set in), and grow
with $W$, i.e. $\a' (W) > 0$.

As for $\b(W,Z)$, we should similarly have a function with a
finite limit $\b_0$ for $W \to 0$; as for its dependence on $W$
and $Z$, we assume it is only through the ratio $X/Z$ of BF
molecules binding to the chromosome and BF molecules in the fluid
environment.

In the following, we choose to deal with a ''minimal'' model, i.e.
consider a linear dependence on $w$ for both these quantities; we
set \beq\label{AB} \a (W)  =  a_0  +  a_1  W \ , \ \ \b (W) = b_0
 +  b_1 (W/Z) \ ; \eeq this implies of course \beq a (w) = a_0 +
a_1 N w \ , \ \ b (w) = b_0 + b_1 (w/z) \ . \eeq

The parameters $(a_0,a_1,b_0,b_1)$ are all positive (we could
always set one of these parameters equal to one by rescaling the
time unit.) We also stress that in view of the discussion by
Nicodemi and Prisco \cite{NP}, we should expect the cooperative
effects enhancing affinity should be predominant over those
depressing the escape rate; this means we should expect $a_1 \gg
|b_1|$.

With these choices, the evolution equations for densities
\eqref{eveq4} read \beq\label{eveq4AB}
\begin{array}{rl}
d x / d t =& (a_0 + a_1 N x ) \, (1 - x - y)
\ - \ [b_0 + (b_1/(1 - N(x+y))) N x] \, x \ , \\
d y / d t =& (a_0 + a_1 N y ) \, (1 - x - y)
\ - \ [b_0 + (b_1/(1
 - N(x+y))) N y] \, y \ .
\end{array} \eeq

We are specially interested in the large $N$ regime. In order to
study this, it is convenient to rescale time (note that
eqs.\eqref{eveq4AB} become singular for $N \to \infty$) and pass
to consider as independent variable \beq \tau \ := \ N t \ . \eeq
We also set \beq \eps \ := \ 1/N \eeq (hence $t = \eps \tau$); the
equations \eqref{eveq4} read then \beq\label{eqseps}
\begin{array}{rl}
d x / d \tau  =& a_1 (1 - x - y) x + \eps \, [a_0 (1 - x - y) -
b_0 x] \ + \ \eps \, b_1 x^2 (x+y - \eps)^{-1} \ , \\
d y / d \tau  =& a_1 (1 - x - y) y + \eps \, [a_0 (1 - x - y) -
b_0 y] \ + \ \eps \, b_1 y^2 (x+y - \eps)^{-1} \ ; \end{array}
\eeq at first order in $\eps$, we have \beq\label{eqslim}
\begin{array}{rl}
d x / d \tau  =& a_1 (1 - x - y) x \ + \ \eps \, [a_0 (1 - x - y)
 \ - \ (b_0 - b_1 x /(x+y)) x ] \ , \\
d y / d \tau  =& a_1 (1 - x - y) y \ + \ \eps \, [a_0 (1 - x - y)
 \ - \ (b_0 - b_1 y /(x+y)) y ] \ . \end{array} \eeq

\section{Different time-scales for the dynamics}
\label{sec:timescales}

We want now to discuss some point concerning the qualitative
behavior of the dynamical system defined by \eqref{eveq4AB}; we
should consider it in the region (invariant under the dynamics) of
$\R^2$ delimited by the coordinate axes and by the line $x+y=1$.

\begin{figure}
\begin{tabular}{|ll|}
\hline
(a) & \includegraphics[width=250pt]{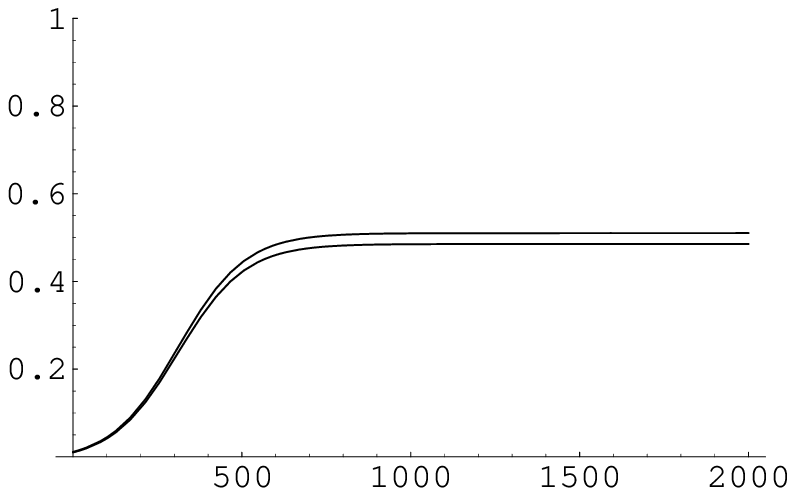} \\
(b) & \includegraphics[width=250pt]{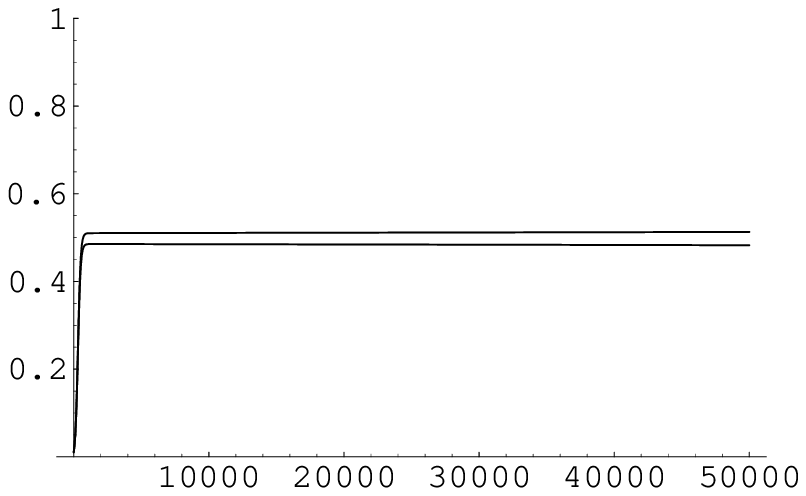} \\
(c) & \includegraphics[width=250pt]{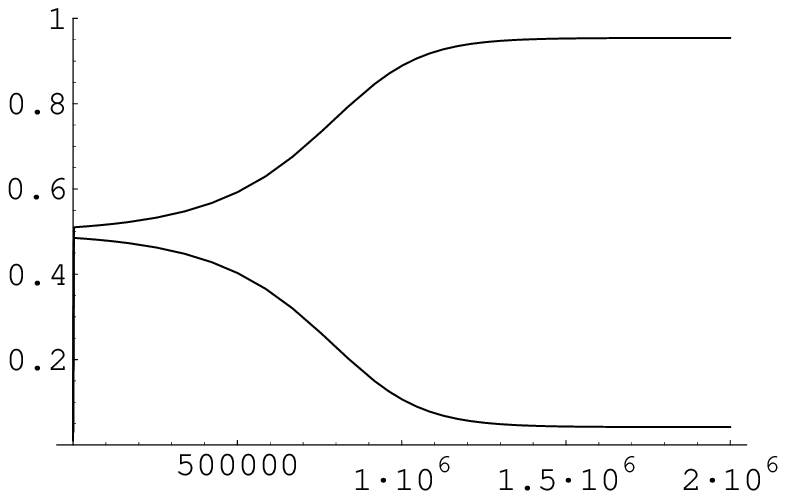} \\
\hline \end{tabular}
  \caption{Evolution of $x(t)$ (upper curves) and $y(t)$ (lower curves)
  as a function of the rescaled time $\tau$, obtained numerically
  integrating the equations \eqref{eqseps}.
  The parameters defining the model have been set as follows:
  $a_0 = 1$, $a_1 = 0.01$, $b_0 = 0.5$, $b_1 = 0.1$; while for $N$ we have
  used $N=10000$, giving $\eps = 0.0001$. Initial conditions are $x(0)= 0.011$,
  $y(0)=0.01$.
  (a): the initial expansion (this plot shows the evolution
  for $0 \le \tau \le 10^3$); note the two densities grow together.
  (b): on a longer timescale one observes how
  the slow phase sets in
  and the densities remain nearly constant over a long time
  (this plot shows the evolution for $\tau \le 5 \cdot 10^4$).
  (c): on a still longer timescale, the transition from the
  saddle point $E_3$ to the stable equilibrium -- in this case,
  $E_2$ -- is clearly visible and happens at an intermediate speed
  (this plot shows the evolution for $\tau \le 2 \cdot 10^6$).}\label{fig1}
\end{figure}

\begin{figure}
\begin{tabular}{|ll|}
\hline
(a) & \includegraphics[width=250pt]{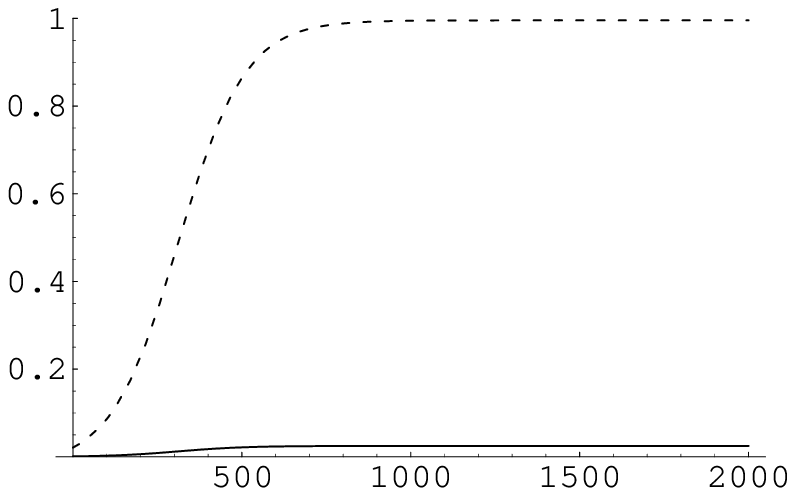} \\
(b) & \includegraphics[width=250pt]{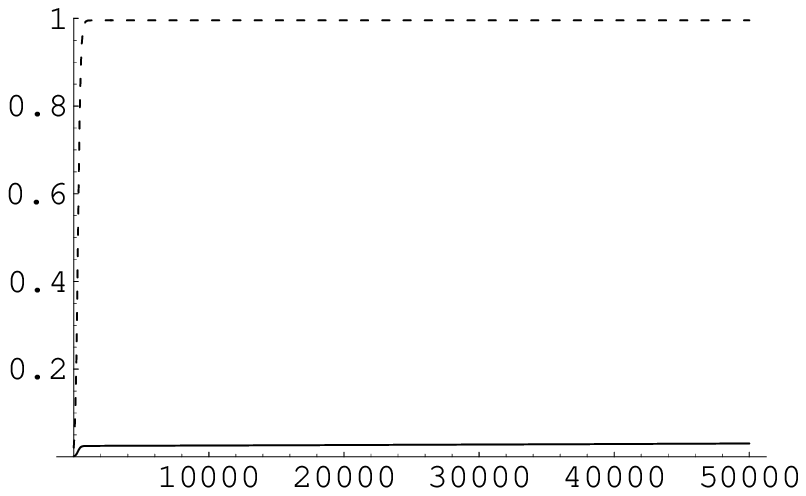} \\
(c) & \includegraphics[width=250pt]{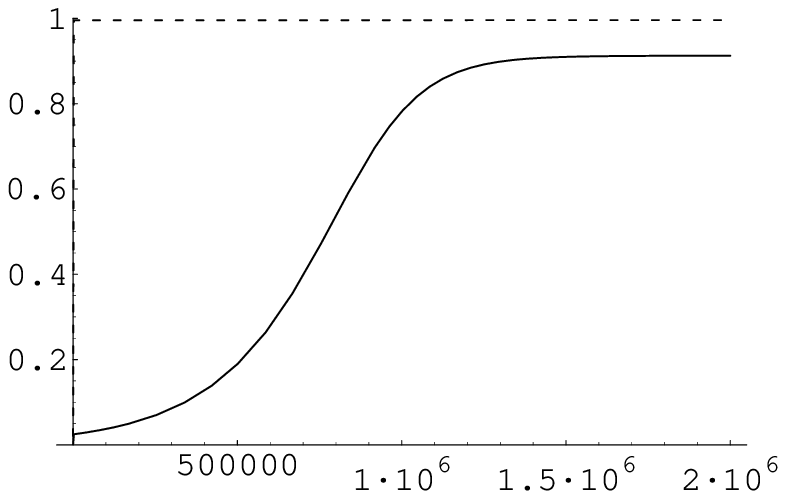} \\
\hline \end{tabular}
  \caption{Evolution of $\xi(t) = x(t)+y(t)$ (dotted curves) and
  $\eta(t)= x(t) - y(t)$ (solid curves) as a function of the
  rescaled time $\tau$, obtained numerically
  integrating the equations \eqref{eqseps}.
  Parameters and initial conditions as in Figure \ref{fig1}.
  (a): in the initial expansion $\xi$ grow abruptly, while $\eta$
  remains near to zero (this plot shows the evolution
  for $0 \le \tau \le 10^3$).
  (b): in the slow phase both $\xi$ and $\eta$
  remain nearly constant over a long time
  (this plot shows the evolution for $\tau \le 5 \cdot 10^4$).
  (c): in the intermediate speed phase, $\xi$ remains nearly constant, while
  $\eta$ undergoes a relatively fast growth
  (this plot shows the evolution for $\tau \le 2 \cdot 10^6$).}\label{fig2}
\end{figure}

\begin{figure}
\begin{tabular}{|ll|}
\hline
(a) & \includegraphics[width=250pt]{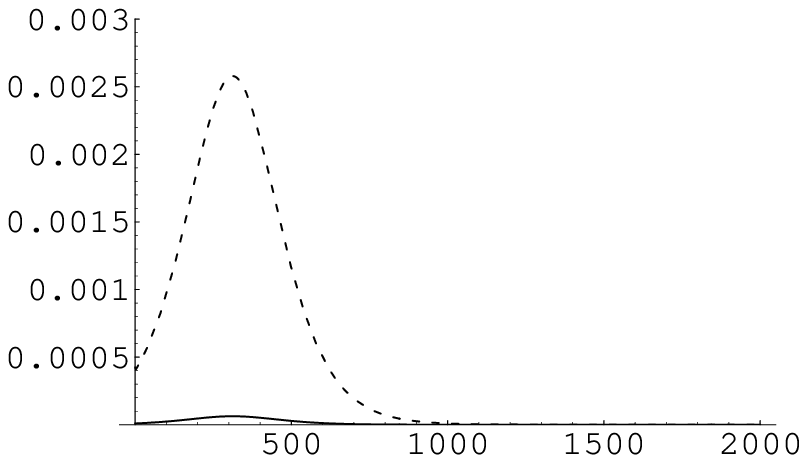} \\
(b) & \includegraphics[width=250pt]{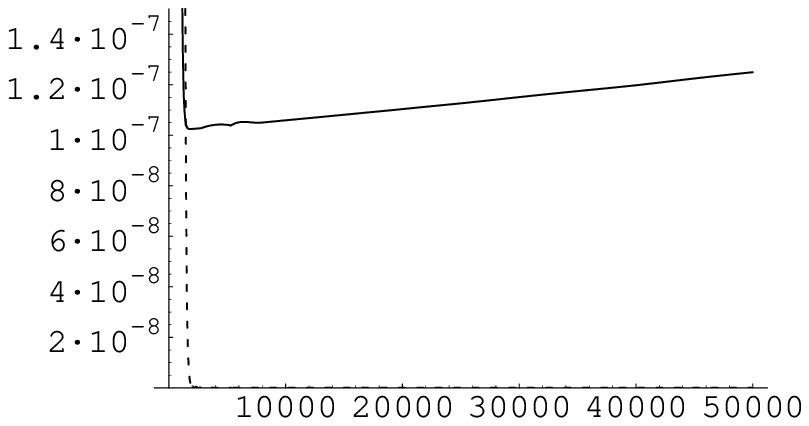} \\
(c) & \includegraphics[width=250pt]{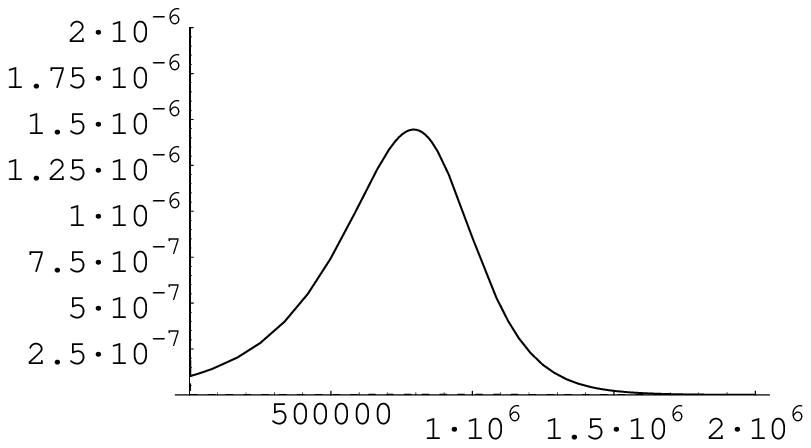} \\
\hline \end{tabular}
  \caption{Evolution of $\xi'(\tau)$ (dotted curves) and
  $\eta'(\tau)$ (solid curves) resulting from the numerical
  integration of the equations \eqref{eqseps}.
  Parameters and initial conditions as in Figure \ref{fig1}.
  (a): in the initial expansion $\xi' (\tau)$ reaches values of order
  $2.5 \cdot 10^{-3}$, while $\eta' (\tau)$ remains smaller than $7 \cdot
  10^{-5}$.
  (b): in the slow phase both speeds are very small, but $\xi' < 10^{-8}$
  while $\eta' \simeq 10^{-7}$.
  (c): in the intermediate speed phase, $\xi'$ remains extremely small (in this
  case we actually have $\xi' \simeq 10^{-9}$), while
  $\eta'$ slightly grows again; in this case $\eta' \simeq 10^{-6}$.}\label{fig3}
\end{figure}

As noted above, the symmetric fixed point $E_3$ is a saddle point
for the dynamics. It is easy to see that its (invariant) stable
manifold \cite{Gle,GuH,Ver} is just the line $y=x$.

Let us consider the situation where at first very few BF molecules
are binding to either one of the X-chromosomes, i.e. $x(0) \approx
0 \approx y(0)$; note that as $x(0) \approx y(0)$, the initial
data are close to the stable manifold for the saddle point $E_3$.

In a first stage, both $x(t)$ and $x(t)$ will grow exponentially
in $\tau$, with exponent $a_1$; this is easily seen from
\eqref{eveq4AB} (assuming $x \ll 1$, $y \ll 1$). More precisely,
near the origin one has $dx/d \tau \approx a_1 x$, $d y / d \tau
\approx a_1 y$; it follows that in this first phase, $x(t) \approx
y(t)$, i.e. the dynamics remains in a neighborhood of the
symmetric line $x=y$.

This fast growth will go on until the system gets near to $E_3$
(note this also means  $1 - (x + y) = O (\eps)$); at this point, a
slower evolution gets in. Once it reaches the vicinity of $E_3$,
the dynamics is dominated by the expansion rate near $E_3$, i.e.
by the positive eigenvalue $\la_2 (E_3) \approx (b_1/2) \eps$, and
evolution takes place at a speed of order $\eps$.

However, at some point (after a time of order $1/\eps$) the system
gets far enough from $E_3$ to have again a dynamics non fully
described by the linearized system identified by $M_3$, and a
somewhat faster -- albeit not as fast as in the first phase --
evolution can take place.

This will lead the system near one of the non-symmetric equilibria
(depending on the initial data), where again the dynamics is
dominated by the eigenvalue which is smaller in modulus; this is
$\la_2 (E_{1,2}) = \eps b_1$, i.e. the system gets again a speed
of order $\eps$, and thus approaches the final equilibrium, as
already discussed, with a rate of order $\eps$.

Summarizing, the evolution of our system will have (assuming
initial conditions are near to zero) four distinct phases:
\par\medskip\noindent
(1) A phase of rapid growth, exponential with expansion rate of
order $a_1$;
\par\noindent
(2) A slow phase (speed of order $\eps$) spent in the vicinity of
the saddle point $E_3$;
\par\noindent
(3) An intermediate (moderate) speed phase, in which the system
travels from a neighborhood of $E_3$ to a neighborhood of either
ones of the asymmetric stable equilibria $E_{1,2}$ with a speed
substantially higher than $\eps$;
\par\noindent
(4) A new slow phase, in which the system approaches exponentially
the stable equilibrium at a rate of order $\eps$.
\medskip

This behavior is clearly shown in Figure \ref{fig1}, where we plot
the numerical solution to the full equations for $x(\tau)$ and
$y(\tau)$ (that is, without truncation to first order in $\eps$)
on different time scales.

In Figure \ref{fig2}, we look at the evolution of the quantities
$\xi (t) := x(t) + y(t)$ and $\eta (t) := x(t) - y(t)$,
representing the total fraction of BF molecules binding to either
one of the X-chromosomes and the symmetry breaking measure
respectively. This again shows clearly that the overall evolution
of the system goes through the different phases enumerated above.

Confirmation is also obtained by Figure \ref{fig3}, where we plot
$d \xi / d \tau$ and $d \eta / d \tau$ on different timescales.

\section{Fluctuations}
\label{sec:noise}

The model equations \eqref{eqseps} or the limit ones
\eqref{eqslim} only deal with evolution of average quantities,
i.e. do not take into account in any way fluctuations. Our present
task is to go beyond this level of description, and take into
account {\it fluctuations} around the average dynamics.

As a first approximation, these can be taken into account by
transforming the equations \eqref{eqseps} or \eqref{eqslim} into
stochastic differential equations \cite{Gue,Oks,vKam} by adding a
noise term (see the Appendix for details). In other words, if the
deterministic equations are $dx/d\tau = f(x,y)$, $d y / d \tau =
g(x,y)$, we will write \beq\label{eqsepsnoise}
\begin{array}{rl}
d x  =& f(x,y) \, d \tau \ + \ \ga \, d \omega (\tau) \ , \\
d y  =& g(x,y) \, d \tau \ + \ \ga \, d \omega (\tau ) \ ,
\end{array} \eeq
 with $\omega (\tau)$ a standard Wiener process and $\ga$ a
parameter describing the size of fluctuations. Note that (as these
are fluctuations for an average) $\ga$ will depend on $N$ and
hence on $\eps$; more precisely, $\ga \simeq \ga_0 \sqrt{\eps}$.
This dependence -- and the estimate $\ga_0 \le \ga_0 =
\sqrt{a_1}/2$ -- can be obtained by a standard procedure.

\begin{figure}
\begin{tabular}{|c|}
\hline
  \includegraphics[width=250pt]{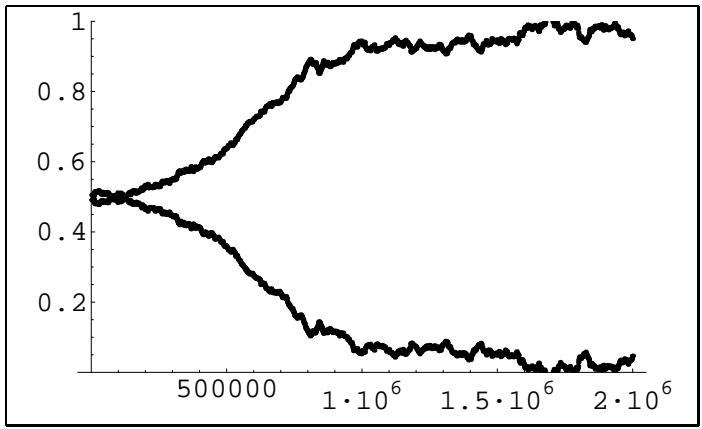} \\
\hline
\end{tabular}
  \caption{Simulation of the stochastic process defined by \eqref{eqslimnoise};
  the parameters of the model have been chosen as in the previous Figures (see
  caption to Figure \ref{fig1} for their values), and we have set $\ga = 10^{-4}$.
  In this case the fluctuation do not seriously modify the dynamics, and the
  discussion based on the deterministic model confirms its validity.}\label{fig4}
\end{figure}

At first order in $\eps$, we have \beq\label{eqslimnoise}
\begin{array}{rl}
d x   =& \[ a_1 (1 - x - y) x + \eps [a_0 (1 - x - y) \right. \\
& \left. - (b_0 - b_1 x /(x+y)) x ] \] d \tau + \sqrt{\eps} \ga_0 d \omega (\tau) \ , \\
d y   =& \[ a_1 (1 - x - y) y + \eps [a_0 (1 - x - y) \right. \\
 & \left. - (b_0 - b_1 y /(x+y)) y ] \] d \tau + \sqrt{\eps} \ga_0 d \omega (\tau)
\ . \end{array} \eeq
 When $1-x-y = O(\eps)$, the noise term could
-- depending on the relation between $\eps$, $a_1$ and $\ga_0$ --
dominate the dynamics. That is, in the slow phase near the saddle
point (see the discussion in section \ref{sec:timescales}) random
motion due to fluctuations could be dominant: if $\ga$ is large
enough, the time spent in a neighborhood $\mathcal{B}$ of $E_3$
will be determined by the escape time needed to exit $\mathcal{B}$
due to fluctuations. This will produce an acceleration of the
dynamics with respect to the purely deterministic one -- and
possibly could cause the system to cross the line $x=y$ and end up
near the other stable equilibrium (this happens in the numerical
simulation plotted in figure \ref{fig4}). On the other hand, if
$\ga$ is small enough then the fluctuations will cause just a
twiggling of trajectories around deterministic ones.

Obviously dominance of fluctuation in the dynamics will take place
once the system gets near -- or very near if $\ga$ is small -- to
the stable equilibrium point $E_1$ or $E_2$. There will be a
metastable equilibrium distribution around $E_{1,2}$ and -- over
extremely long times -- large fluctuations could drive the system
near to the other stable equilibrium; the equilibrium distribution
and the time needed for large fluctuations to appear can be
estimated via standard stochastic processes techniques
\cite{vKam}.

In Figure \ref{fig4} we show a numerical simulation of (the
stochastic process defined by) the stochastic differential
equations \eqref{eqslimnoise}.

\section{Conclusions}

We have provided a mean-field formulation of the model recently
proposed by Nicodemi and Prisco \cite{NP} for the X-chromosome
inactivation in mammals \cite{LKP,CYZB,CTTB,NTL,NCB,Xclose}.

Like their model, the version proposed here explains
$X$-chromosome inactivation as the result of a {\it spontaneous
symmetry breaking} in the dynamics of blocking factors molecules
binding to the X-chromosome inactivation centers (Xic). While
their approach was based on a single-agent model, in our version
the considered equations deal directly with average quantities.

After discussing a general formulation of the model, we considered
a minimal version, in which the affinity grows linearly with the
number of BF molecules already binding to the Xic and the rate of
escape of BF molecules increases linearly on the ratio between
density of BF molecules binding to the X-chromosomes and of BF
molecules floating in the fluid environment with the same number.
Our model contains four positive parameters, related to affinity
and escape probability for the BF molecules.

We showed that when $b_1 < b_0$, non-symmetric equilibria exist;
under the same condition they are stable (while symmetric
equilibria are unstable). Thus in such case we have a {\it
spontaneous symmetry breaking}, i.e. non-symmetric equilibria
describe the asymptotic state of the system. We were also able to
estimate the rate of approach to equilibrium as a function of the
parameters describing the system and the total BF population.

A qualitative discussion showed that when the initial situation
corresponds to most of the BF molecules floating in the fluid, one
should expect the dynamics to undergo different phases. We ran
some numerical simulations (one of these is shown in the Figures)
confirming the conclusions reached by the qualitative discussion
mentioned above. We also briefly discussed how fluctuations can be
taken into account within our scheme.

Finally, we would like to stress that our model differs from that
of Nicodemi and Prisco only in the mathematical description it
considers, while the physical basis is the same. Thus, we are just
providing an alternative description -- in terms of dynamical
systems and ODEs rather than of stochastic processes -- for the
Physics described by Nicodemi and Prisco in their paper \cite{NP}.

This mathematically simpler description allows to make more
detailed predictions, in particular about different timescales in
the dynamics; these were fully confirmed by numerical simulations.

Our model, like the one of Nicodemi and Prisco \cite{NP}, did not
take into account co-localization; work is in progress to take
this into account within the present description of
X-inactivation.

\section*{Appendix}
\label{sec:noise:technical}

In this Appendix we will briefly discuss the passage from the
physical model of section \ref{sec:model} to the stochastic
differential equations \eqref{eqsepsnoise} and
\eqref{eqslimnoise}.

In order to take into account full detail of fluctuations in our
model, we should consider the flows $\Phi_{ij}$ of particles from
state $i$ to state $j$. Denoting by $\nu_{ij} (\de t)$ the number
of particles passing from state $i$ to state $j$ in a time
interval of length $de t$, we have of course \beq \Phi_{ij} =
\nu_{ij} - \nu_{ji} \ . \eeq The transition probabilities $p_{ij}$
for a single particle are given in section \ref{sec:mod:TP}; the
$\nu_{ij}$ will thus be Poisson distributed with parameter
$\la_{ij} = p_{ij} n_i$, where $n_i$ is the population of state
$i$ at the beginning of the time interval.

Needless to say, if we look just at expectation values we obtain
again \eqref{eveq1}; however, our present task is to go beyond
this level of description, and take into account first and higher
order momenta of the Poisson distribution -- that is, in physical
terms, {\it fluctuations} around the average dynamics.

Let us fix our attention, for the sake of concreteness, on the
population $X$ (or density $x = X/N$) of the state 1 and its
variation $\de X$ (or $\de x$) in a time interval of length $\de
t$; we also write $Z$ and $z$ rather than expressing these
quantities in terms of $(X,Y)$ and $(x,y)$, for ease of notation.

We obviously have $de X = \nu_{31} - \nu_{13}$. Recall that
$\nu_{31}$ is Poisson distributed with parameter $\la = \a (X) Z
\de t$ (i.e. $\mathcal{P} (\nu_{31} = k) = \la^k e^{-\la}/k!$),
and $\nu_{13}$ is also Poisson distributed but with parameter $\mu
= \b (X,Z) X$. Thus $\Phi_{31}$ is Poisson with mean $\la - \mu$
and variance $\s^2_{31} = (\la + \mu )$. Thus for $X$ we have the
stochastic differential equation $ d X = (\la - \mu) d t + \s d
\om(t)$; in the case of our minimal model this reads
 \beq
 d X = [(a_0 + a_1 X) Z - (b_0 + b_1 X/Z)X] d t
 + \sqrt{(a_0 + a_1 X) Z - (b_0 + b_1 X/Z) X} d \om(t) \ . \eeq
Passing to density variables this reads $d x = N^{-1} [(a_0 + a_1
N x) N z - (b_0 + b_1 x/z) N x] d t  + N^{-1}[(a_0 + a_1 N x) -
(b_0 + b_1 x/z) N x ]^{1/2} d \om(t)$; we should also pass to the
rescaled time variable $\tau = N t$, so that $dt = (1/N)(d \tau )$
and $d w (t) = (1 / \sqrt{N}) d \om (\tau )$.

In this way we get, using again $N^{-1} = \eps$,
 \beq
 d x = [a_1 x z + \eps (a_0 z - (b_0 + b_1 x/z) x ]  d \tau
  + \sqrt{a_1 x z + \eps (a_0 z  - (b_0 + b_1 x/z) x}
 \sqrt{\eps} \, d \om(\tau) \ . \eeq
 We thus have a noise of intensity $\ga = \^\ga_0 \sqrt{\eps}$; note
 that at order zero in $\eps$, $\^\ga_0 = \sqrt{a_1 x z}$; as the
 sum of $x$ and $z$ cannot be higher than one, this can be
 estimated by $\^\ga_0 \le \ga_0 = \sqrt{a_1}/2$.

\label{lastpage}
\end{document}